
\input phyzzx
\def\tcm{topological coset model}
\def\tca{``topological coset model"}
\def\ala{affine Lie algebra\ }
\def\SOS {$SL(2,R)\over SL(2,R)$}
\def\SOU {$SL(2,R)\over U(1)$}
\def\co{cohomology\ }
\def\G{${G\over G}\ $}

\def\ch#1#2{\chi_{#1}^{#2}}

\def\rh#1#2{\rho_{#1}^{#2}}

\def\c#1#2{\chi_{#1}^{#2}}

\def\r#1#2{\rho_{#1}^{#2}}
\def\jt#1#2{{J^{(tot)}}_{#1}^{#2}}

\def\t{\tilde}
 \def\pa{\partial}
\def \f{f^a_{bc}}
\def \fcr{\f \c b {} \r c {} }
\def \jg{J^{(gh)}}

\def \pa{\partial}

\def\S#1{$SL({#1},R)$}
\def\A {$A_{N-1}^{(1)}$}

\def\G{${G\over G}\ $}
\def\GH{${G\over H}\ $}
\def\tGH{twisted\ ${G\over H}\ $}

\def\TKS{twisted Kazama-Suzuki\ }
\def \vh{\vec{\t {\cal H}}}
 \def\t{\tilde}
 \def\pa{\partial}
\def\bpa{\bar\partial}

\def\bP{\beta^+}
\def\bM{\beta^-}

\def\gP{\gamma^{+}}
\def\gM{\gamma^{-}}

\def\pP{\phi^+}
\def\pM{\phi^-}

\def\dpP{\partial\phi^+}
\def\dpM{\partial\phi^-}

\def\cP{\chi^+}
\def\cz{\chi^0}
\def\cM{\chi^-}

\def\rP{\rho^+}
\def\rz{\rho^0}
\def\rM{\rho^-}

\def\sq2{{\sqrt 2}}

\def\d{\partial}

\def\QB{Q_{BRST}}
\def\QBr{{\QB}^{(rel)}}
\def\Qt{Q_{tr}^{(rel)}}
\def\intz{\oint{dz\over{2\pi i}}}

\def\cmp#1{{\it Comm. Math. Phys.} {\bf #1}}

\def\pl#1{{\it Phys. Lett.} {\bf #1B}}

\def\prd#1{{\it Phys. Rev.} {\bf D#1}}

\def\np#1{{\it Nucl. Phys.} {\bf B#1}}

\def\jmath#1{{\it J. Math. Phys.} {\bf #1}}

\REF\Wym{E. Witten, \cmp {117} (1988) 353.}
\REF\dvv{R. Dijkgraaf, E. Verlinde, and H. Verlinde,
\np {352} (1991) 59.}
\REF\MS{D. Montano, J. Sonnenschein ,  \np {324} (1989) 348,
J. Sonnenschein \prd {42} (1990) 2080.}
\REF\Wgg{E. Witten, ``On Holomorphic Factorization of WZW and Coset
Models" IASSNS-91-25.  }
\REF\SY{M. Spiegelglas and S. Yankielowicz
\np {393},(1993) 301.}

\REF\BaRS{K. Bardacki, E. Rabinovici, and B. Serin \np {299} (1988)
151.}  \REF\GK{
  K. Gawedzki and A. Kupianen , \pl {215} (1988) 119, \np
     {320} (1989)649.}
\REF\KS{D. Karabali and H. J. Schnitzer, \np {329} (1990) 625.}
\REF\usss  {O. Aharony,O. Ganor  J. Sonnenschein and S.
Yankielowicz , \np {399} (1993) 560.}
\REF\KaSu{ Y. Kazama and H. Suzuki, \np {321} (1989).}
\REF\IsiRa{J. Isidro and A. V. Ramallo `` gl(N,N) current algebras and TFTs"
US-FT-3/93.}
\REF\BMP{P. Bouwknegt, J. McCarthy and  K. Pilch \pl {234} (1990) 297,
\cmp {131} (1990) 125;Cern Preprint
TH-6162/91}
\REF\us  {O. Aharony,O. Ganor N. Sochen  J. Sonnenschein and S.
Yankielowicz ,\np {399} (1993) 527.}
\REF\uss  { O. Aharony, J. Sonnenschein and S.
Yankielowicz ,   \pl {289} (1992) 309.}
\REF\BF{D. Bernard  and G. Felder \cmp {127}  (1991)  145.}
\REF\FaLu{ Fateev and Lukeanov
{\it Int. Jour. of Mod. Phys. }{\bf A31} (1988)
507.}
\REF\BerK{M. Bershadsky and I. Klebanov \np {360} (1991) 559.}
 \REF\LZ{B. Lian and G. Zuckerman \pl   {254}  (1991) 417.}
 \REF\ussss  {O. Aharony,O. Ganor  J. Sonnenschein and S.
Yankielowicz , \pl {305},(1993) 35-42.}
\REF\Wgr{ E. Witten  \np {377} (1992) 55.}
\REF\MuVa{ S. Mukhi and C. Vafa `` Two Dimensional Black hole as a
Topological Coset model of $c=1$ String Theory",HUTP-93/A002,
TIFR/TH/93-01.}

\rightline{TAUP- 2098-93}
\date{Oct. 1993}
\titlepage
\vskip 1cm
\title{ Non-Critical String Models as  Topological Coset models. }
\author { J. Sonnenschein and S. Yankielowicz
\footnote{*} {Work supported in part by the US-Israel Binational
Science Foundation and the Israel Academy of Sciences.}}
 \address{ School of Physics and
Astronomy\break Beverly and Raymond Sackler \break Faculty of Exact
Sciences\break Ramat Aviv Tel-Aviv, 69978, Israel}
\abstract{
The  topological coset model  appraoch to non-critical
string models is
summarized. The action of a  topological twisted
${G\over H}$ coset model
($rank\ H = rank\ G$) is written down.
A ``topological coset algebra" is derived  and compared with
the algebraic structure of the $N=2$ twisted models.
 The cohomology on a free field  Fock
space  as well as on the space of  irreducible representation of the
``matter" \ala are extracted.  We compare the results of the
$A_1^{(N-1)}$ at level $k={p\over
q}-N$ with  those of  $(p,q)$ $W_N$ strings.

A \SOS model which corresponds to the $c=1$ is written down. A
similarity transformation on the BRST charge enables us to extract
 the full BRST cohomoloy. One to one correspondence  between the
physical states of the $c=1$ and the corresponding coset model is
found.

Talk presented  at ``Strings' 93", May 93 Berekely.}
\overfullrule=0pt
\baselineskip=16pt


\section{ Introduction}
Topological quantum field theoreis\refmark\Wym (TQFTs)
were recently  found to be a
very useful tool in the study of string theories.
Non-critical string models or  2D gravitational models serve as a
laboratory to explore  new domains of string theories
such as  the non-pertubative behaviour.
In this talk we summarize the attemp to analayze the $c\leq 1$
models and their $W_N$ generalizations as TQFTs.

A TQFT is a QFT in
which all the  observables, namely, all correlators of ``physical
operators", are invariant under any arbitrary deformation of
 $g_{\alpha\beta}$ the metric of the underlying space-time.
 Given a set of physcial operators $F_i[\Phi^a (x_i)]$ which are functional
of  the  fields $\Phi^a (x)$, $a=1,...p$ of the theory and which are invariant
under the symmetries of the theory, then the theory is topological iff
$$\delta_{g_{\alpha\beta}}<\prod_i F_i[\Phi^a (x)]>=0.\eqn\mishtop$$
In particular this definition implies that all correlators are independent
of distances between the operators.
In a theory where  the energy-momentum  tensor
$T_{\alpha\beta}$ is exact under a BRST-like  symmetry operator $Q$, namely,
$T_{\alpha\beta}=\{Q, G_{\alpha\beta}\}$,
 all physical
operators  are in the  cohomology of $Q$
 property \mishtop\ is obeyed. Thus, it is a TQFT.

A  two-dimensional theory  which is  a TQFT as well as conformal,
 is a topological conformal field theory
(TCFT). The algebraic structure of these models is characterized
by the fact that  $T$ and the $Q$ are  both BRST exact.
However, there is no unique structure which is common to all TCFTs.
In fact it is shown here that the ``topological coset algebra"
differ from the algebra of the twisted $N=2$ models.\refmark\dvv

 Every conformal field theroy coupled to 2-D gravity is
obviously a TCFT
after integrating over all metric degrees of freedom.
We now define the concept of a ``topological model" which  is a
TQFT without the introduction and integration over the metric.
In more than two
dimensions   examples of such models are the Chern-Simons theory and the four
dimensional theory which corresponds to  the Donaldson
invariants.\refmark\Wym In two dimensions an example is the theory of flat
gauge
connections.\refmark\MS We are now ready to introduce the  notion of
``topological coset models" which are  gagued WZW models that are also
topological models.\refmark{\Wgg,\SY}

The paper is organized as follows. In section 2 we write
down           the  quantum action of a  topological  \G and \tGH
coset models  as a sum of ``decoupled" matter, gauge and ghost
sectors.  The algebraic structure is derived in section 3
and it is found that
a larger algebra then that of the TCFT\refmark\dvv algebra.  Section
4. is devoted to   a breif remineder of the cohomology on a free
field  Fock space  as well as on an irreducible representation of
the ``matter" \ala .  In section 5 a comparison between
the physical states of \tcm s and the corresponding
string models is made. We discuss
the realtion between gauge holonomies and the twist needed to acheive
the correspondence with the gravitationsl model.  We then write down
a \tcm\  which describes the $c=1$ string model. A special treatment
of the BRST charge using a  double similarity tranformation is used
to exactract the space of physical states.  The latter are then
compared with the ground ring and other states of the two
dimensional string theory.

\section{ The quantum action}
The   \tcm s
 are constructed by gauging   an anomaly free diagonal  subgroup $ H \in G$
group of a level $k$   $WZW$ model.  The \G models are defined by
$H=G$ whereas the general \tGH models  require that $rank\ H=\ rank\
G$. In the latter case the gauged action is a twisted
supersymmetric  $G$-WZW model, namely,   the usual \GH model with an
extra  set of $(1,0)$ anti-commuting ghosts where the dimension one
fields take their values in the positive roots of \GH and the
dimension zero fields in the negative ones. A great part of the
discussion that follows applies to general compact groups,  however,
we will mainly
concentrate on $G=SU(N)$ and the non-compact $SL(N)$ groups.
The latter are needed since
  we will be interested
also in fractional levels.   We proceed to derive the quantum action of the \G
and the \tGH models.

\subsection{ The \G model}

 The  classical action  of the \G model takes the form
$$ S_k(g,A,\bar A ) = S_{k}(g) +i{k\over 2\pi}
\int_{\Sigma}d^2 z Tr(g^{-1}\pa g \bar A +  g\bar \pa g^{-1} A -
\bar A g^{-1}A  g  + A \bar A  )\eqn\mishwzw$$
where $g\in G$ and $S_k(g)$ is the WZW action at level $k$.
An essential step in the analysis of this  action is a ``decoupling" of the
matter and gauge degrees of freedom.  This can be acheived by
rewriting the gauge fields in terms of group elements.
In  the case of
 a topologically trivial $\Sigma$ the gauge  field can be parametrized as
follows
$A =ih^{-1}\pa  h ,  \bar  A =i{\bar h}\pa  {\bar h}^{-1}$ where
$h(z)\in G^c$. The action then\refmark{\BaRS,\GK,\KS}
reads
$$S_k(g,A) =S_k(g) -S_k(h \bar h) \eqn\mishwzwh$$
The Jacobian  of
the change of variables introduces a   dimension $(1,0)$ system of
anticommuting ghosts $\chi$ and $\rho$ in the adjoint representation of the
group. The quantum action thus
 takes the form  of
$$S_k(g,h,\rho,\chi) =S_k(g) -S_{k}(h\bar h) -i\int d^2z Tr[\rho\bar
D \bar\chi+ c.c] \eqn\mishwzwh$$
 where
$D\chi=\pa\chi-i[A,\chi]$.  This action involves
an interaction term of the
form $Tr_H(\bar\rho [h^{-1}\pa h,\bar\chi])$ and a similar term for
$\rho,\chi$. By performing a chiral rotation $\bar\rho\rightarrow
h^{-1}\bar\rho h$ and $\bar\chi\rightarrow h^{-1}\bar\chi h$ with
$\rho\rightarrow\bar h\rho\bar h^{-1}$ and $\chi\rightarrow\bar
h\chi\bar h^{-1}$, one achieves a decoupling
of the whole ghost system. The price of that is an additional
$S_{-2C_G}(h\bar h)$ term in the action resulting form the corresponding
anomaly.  $C_G$ is the second Casimir of the adjoint
representation.
This result can be derived by using a non-abelian bosonization of the
ghost system.\refmark{\usss}  The  final quantum action after gauge
fixing $\bar h=1$
 $$S_k(g,h,\rho,\chi) =S_k(g) -S_{-(k+2C_G)}(h) -i\int d^2z
Tr[\rho\bar \pa \chi
+ c.c], \eqn\mishwzwh$$
indeed, has the structure of three ``decoupled" \ala actions.

\subsection{ The \tGH model}

The  classical action of the  \tGH model is given by
 $$ \eqalign{S_{(tKS)} &= S_k(g,A,\bar A) + S_{(gh)}^{G\over H}\cr
  S_k(g,A,\bar A)  &= S_{k}(g)
-{k\over 2\pi} \int_{\Sigma}d^2 z Tr_G[g^{-1}\pa g \bar A_{\bar z} +  g\bar \pa
g^{-1} A  - \bar A g^{-1}A  g  + A \bar A  ]\cr
S^{G\over H}_{gh}&={i\over {2\pi}}\int d^2z \sum_{\alpha\in{ G\over
H}}[\rho^{+\alpha}(\bar D \chi )^{-\alpha}+ \bar\rho^{+\alpha}( D
\bar\chi)^{-\alpha}]\cr}
\eqn\mishwzw$$
Using the same parametrization as for the \G case, one finds after
inserting the Jacobian and gauge fixing
 the following action
$$S_k(g,A) =S_k(hg\bar h) -S_k(h\bar h) +{i\over {2\pi}}\int d^2z
Tr_H[\rho\bar D \chi + \bar\rho D \bar\chi]. \eqn\mishwzwh$$
The \TKS\refmark\KaSu action is
given by
 $$ \eqalign{S_{(tKS)} &= S_k(g) - \sum_{I=1}^n S_k(h^{(I)}) -
{k\over 4\pi  }\int d^2 z  \sum_{s=1}^r \pa {\cal H}_s\bpa {\cal H}_s\cr
  &+{i\over {2\pi}}\int d^2zTr_H[\rho\bar D \chi + \bar\rho \pa\bar\chi]
+{i\over {2\pi}}\int d^2z \sum_{\alpha\in{ G\over
H}}[\rho^{+\alpha}( \bpa \chi )^{-\alpha}+ \bar\rho^{+\alpha}( D
\bar\chi)^{-\alpha}]\cr}
\eqn\mishwzwHi$$
for
   $G=SU(N)$ and
$H=SU(N_1)\times ...\times SU(N_n)\times U(1)^r$ with $r =N-1-\sum_{I=1}^n
N_I+n$, where  the gauge fields $A$ take the form of $A=
i\sum_{I=1}^n {h^{(I)}}^{(-1)} \pa h^{(I)} +i\sum_{s=1}^r \pa {\cal
H}_s$.

After chiral rotating the ghost fields  the action takes the
following form
   $$ \eqalign{S_k =&S_k(g)
 + \sum_{I=1}^nS_{-(k +C_G +C_{H^{(I)}}) }(h^{(I)})+\cr
 &{1\over 2\pi }\int
d^2z[\sum_{s=1}^r \pa {\cal H}_s\bpa {\cal H}_s
 + i\sqrt{2\over
k+C_G}(\vec\rho_G-\vec\rho_H)\cdot \vh R] \cr
&+{i\over{2\pi}}\int d^2z Tr_H[\bar\rho \pa \chi + \rho\bar\pa\chi]
+{i\over{2\pi}}\int d^2z \sum_{\alpha\in{ G\over
H}}[\rho^{+\alpha}(\bar \pa \chi )^{-\alpha}+ \bar\rho^{+\alpha}( \pa
\bar\chi)^{-\alpha}] \cr}
 \eqn\mishwzwhh$$
where we have  normalized the $\vh$ fields to be free  bosons, and
$\vec\rho_G$ and $\vec\rho_H$ are half the sums of the
positive roots of $G$ and $H$ respectively. The action is composed of three
decoupled sectors: the matter sector, the gauge sector and the ghost sector
involving  ghosts in $H$ and \GH.

\section{ The Topological coset algebra}
An important class of TCFT is the twisted $N=2$ superconformal
field theories. These models are characterized by the ``TCFT
algebra"\refmark\dvv given by
 $$\eqalign{ T(z) =&\{ Q, G(z)\} \ \ \ \ \ Q(z) =[ Q ,
j^\#(z)]\cr
 &\{ Q, Q(z)\}=0\ \ \ \ \    \{ G, G(z)\}=0 \cr}
\eqn\mishalgebra$$
where  the holomorphic currents, $T(z)$, $j^\#(z)$, $Q(z)$
and $G(z)$ are the energy-momentum ternsor,  a $U(1)$
current,   a BRST-like current and  an
anti-commuting dimension two current,respectively. Obviously these
holomorphic currents have also their anti-holomorphic partners with
the same algebra.
 Let us now examine whether the \tcm\  share the same
algebraic structure.
\subsection{ The \G models}

Let us first examine the $G$ \ala.
There are three sets of holomorphic $G$ transformations which leave \mishwzwh\
invariant $\delta_J g=i[\epsilon(z), g] $   $\delta_I h=i[\epsilon(z), h] $
and $\delta_{J^{(gh)}} \chi^a = i\f\epsilon^b \chi^c$ ; $\delta_{J^{(gh)}}
\rho^a
= -i\f\epsilon^b \rho^c$ with $\epsilon$ in the algebra of $G$. The
corresponding currents $J^a$, $I^a$ and ${\jg}^a=\fcr$ satisfy
the $G$ \ala
 at levels $k$,$-(k+2c_G)$ and $2c_G$
respectively.
Out of all possible linear combinations of these cuurent the
following one is very useful,
$$\jt {} {a}   =J^a +I^a +{\jg}^a= J^a +I^a +i\fcr . \eqn\mishbJ$$
It obeys an \ala at level
$$ k^{(tot)} = k  -(k+2c_G) + 2c_G =0. \eqn\mishbk$$
The energy-momentum  tensor  $T(z)$ is a sum of  Sugawara terms of the $J^a$
and  $I^a$ currents and the usual contribution of a $(1,0)$  ghost system,
namely\refmark{\BaRS,\GK,\KS}
$$T(z) = {1\over k+c_G }:J^a J^a: -
{1\over k+c_G }:I^a I^a: +\r  {} a \pa \c   {} a. \eqn\mishbT$$
The corresponding  Virasoro central  charge  vanishes
$$ c^{(tot)} = {k d_G\over k+c_G}  -{(k+2c_G) d_G\over -(k+2c_G)+c_G}
-2d_G =0 \eqn\mishbk$$
This last property is an   indication that the \G model is a
TCFT.

The $Q(z)$ cuurent has an obvious realization. It is just the BRST
current which emerges from the gauge fixing of the original gauged
action. It is then also easy to find its dimension two partner G
$$ \eqalign{ Q(z)=&\c a {} [J^a + I^a + \half {\jg}^a]  \cr
G=&{1\over k +c_G}\r a {} [J^a - I^a ].  \cr}\eqn\mishbGJ$$
It is straightforward to check that
  $T$ is BRST exact,
 $ T(z) =\{ Q, G(z)\} $.
The BRST current itself is also BRST exact with respect to the
$U(1)$ ghost number current $j^\#=\chi^a\rho_a$,
 $ Q(z) =[ Q, j^\#(z)] $. By its construction the BRST charge is
nilpotent so one may conclude that indeed the \tcm\  share the TCFT
algebra.  In fact the algebraic structure is different since $G $
defined above
is not  a nilpotent operator ( apart from the case of $U(1)$). Instead one
finds
$\{ G, G(z)\} \equiv W(z)= {1\over 4C_G}
f_{abc}{J^{(tot)}}^a\rho^a\rho^b + \pa\rho^a\rho_a$.
To close the algebra\refmark\IsiRa one has to introduce one additional
anticommuting current of dimemsion three $U={1\over 12
C_G}f_{abc}{J^{(tot)}}\rho^a\rho^b$ such that the full ``topological coset
algebra" is given by
 $$\eqalign{ \t T(z) =\{ Q, G(z)\} \ \ \   &Q(z) =\{ Q ,
j^\#(z)\}\cr
\{ Q, Q(z)\} =0 \ \ \ \ \ &W(z)\equiv\{ G, G(z)\}\cr
\t W(z) =\{ Q, U(z)\} \ \ \ &[W, W(z)]= 0.\cr}
\eqn\mishtcalgebra$$

\subsection{ The \tGH models}

There are two twisted $N=2$ symmetry algebras in the  \tGH
models. One of them which emerges from the $H$ gauge fixing part, is
 of the ``topological coset algebra" type whereas the other,
which corresponds to the fact that the origianal action is a
 twisted $N=2$ WZW model, is a ``TCFT algebra". For the derivaition
of the ``physical states" the relevant algebra is a  ``topological
coset algebra" which is a direct sum of the
 the two.
Let us start with the $H$  sector. The following set of
currents   obey the \tca\  of eqn. \mishtcalgebra:

$$ \eqalign{T^H(z) &= {1\over2( k+c_G )} g_{ab}:(J^a +J^a_{G\over
H})(J^b+J^b_{G\over
H}):-{1\over 2(k+c_G )}g_{ab}:I^a I^b:
\cr&-{\sqrt{2}\over k+C_G}(\vec\rho_G-\vec\rho_H)\cdot\pa(\vec J +\vec I +\vec
J_{G\over H})
 + g_{ab}\r  {} a \pa \c   {} b\cr
J^\#(z)&= \r {} a \c a {}\cr
Q(z)=&g_{ab}\c  {} a[J^b + I^b + J_{G\over H}^b + \half {\jg}^b_H]
\cr =&g_{ab}\c {} a[J^b + I^b + {i\over 2}f^b_{cd}\r {} c \c {} d  + {i}
f^{b}_{\gamma,-\beta}\rho^\gamma\chi^{-\beta}]\cr
G^H=&{g_{ab}\over 2(k +c_G)}\r  {} a [J^b - I^b +
if^{b}_{\gamma,-\beta}\rho^\gamma\chi^{-\beta}]
-{\sqrt{2}\over
k+C_G}(\vec\rho_G-\vec\rho_H)\cdot\pa\vec\rho
  \cr}\eqn\mishbGJ$$
where
$\vec J$, $\vec I$ and $\vec J_{G\over H}$ are the Cartan-subalgebra currents
given in the basis in which $[J^i_n, J^j_m] = k n \delta ^{ij}\delta_{m+n}$.

The ${G\over H}$ algebra on the other hand involves the following
currnts
$$\eqalign{   T^{G\over H}(z) &= {1\over 2(k+c_G )}g_{\t a\t b}:J^{\t a} J^{\t
b}: -{1\over2(
k+c_G )} g_{ab}:(J^a +J^a_{G\over H})(J^b+J^b_{G\over H}): \cr
&+{\sqrt{2}\over
k+C_G}(\vec\rho_G-\vec\rho_H)\cdot\pa(\vec J  +\vec J_{G\over H})
+\sum_{\alpha\in{ G\over H}}\rho^{+\alpha}( \pa \chi )^{-\alpha}
 \cr
J^\#(z)&= \r {} {+\alpha} \c {-\alpha} {}\cr
Q^{G\over H}&=
\sum_{\alpha\beta\gamma\in {G\over H}}\c {} {-\alpha}(J^\alpha + {i\over 2}
f^{\alpha}_{\gamma,-\beta}\rho^\gamma\chi^{-\beta})\cr
G^{G\over H}&={1\over k+C_G}\sum_{\alpha\beta\gamma\in {G\over H}}\r
{} {\alpha}(J^{-\alpha} + {i\over 2}
f^{-\alpha}_{\gamma-\beta}\rho^\gamma\chi^{-\beta}).\cr }\eqn\mishQG$$
where $\t a$ and $\t b$ go over the adjoint of  $G$.
The combined algebra is based on the set of generators
which have the form $A(z) =A^H(z) + A^{G\over H}(z)$.
The combined energy momentum tensor aquires now the simplified form

$$ \eqalign{ T(z) &= {1\over2( k+c_G )} g_{\t a\t b}:J^{\t a} J^{\t b}: -
{1\over 2(k+c_G )}g_{ab}:I^a I^b: + g_{ab}\r  {} a \pa \c   {} b
\cr &-{\sqrt{2}\over k+C_G}(\vec\rho_G-\vec\rho_H)\pa\vec I
+ \sum_{\alpha\in{ G\over
H}}\rho^{+\alpha}( \pa \chi )^{-\alpha}\cr}\eqn\mishTT$$
with a total Virasoro central  charge is found to be
$$ c= {kd_G\over k+C_G} +\sum_{I=1}^n {(k+C_G+C_{H^I})d_{H^I}\over k+C_G} +r
-2d_H -(d_G-d_H) +6[\sqrt{2\over k+C_G}(\vec\rho_G-\vec\rho_H)]^2 =
0\eqn\mishcz$$ where we have  used, assuming $G$ is a simply laced group,
 the relations  $12\rho_G^2 =d_GC_G$, and
$\vec\rho_H\cdot (\vec\rho_G-\vec\rho_H)=0$.

\section { BRST cohomology   and  Physical states }
Next we  proceed to extract  the space of  physical states of the model.
We take as our definition of a physical state a state
 in  the  \co of  $ Q= Q^{(BRST)} +Q^{G\over H}$, namely, $|phys>\in
H^*(Q)$.
The computation of the cohomology is based on  a spectral sequence
decomposition approach.\refmark\BMP
The extraction of the physical states was worked out
in detail in refs. [\usss,\us,\uss]. Here we will mention one
feature of this method and the final results.
The method is based on `` Wakimoto bosonization" of   the matter ($J$) and
``gauge" ($I$) currents.\refmark{\us}
There are two possible bosonizations  which are related
by an automorphism,
for instance in the $SL(2,R)$ case,   $J^+ \leftrightarrow J^-, J^0
\leftrightarrow -J^0$. Denoting the two parametrizations by $+$ and
$-$ one has the two  options $(+,+)$ and $(+,-)$ for the $(J,I)$
system. In
[\usss,\us,\uss] we  have used a $(+,-)$ scheme since only in this
way a convenient
 grading decomposition  is possible.  Eventhough this bosonization
 lacks an  $SL(2,R)$ invariant vacuum,  after projecting to the space of
irreducible
representations of \ala\refmark\BF\ the appropriate vacuum
invariance is  restored.
The full \co on the Fock space of free fields   was found to be
$$H (Q) \simeq H^{rel} (Q)
\oplus\sum_{\{k_1,...,k_l\}}\c 0 {k_1}... \c 0 {k_l} H^{rel} (Q)
\eqn\mishabsco$$
where  the sum  is over $k_1,...,k_{l}$ namely  all possible
subsets of $1,...,N-1$.

The relative \co, $H^{rel} (Q)$ is the \co on the space of
vanishing zero modes of the components of $\rho$ in the Cartan
sub-alegebra of $H$ which is given by
 $$ H^{rel} (Q) = \{\prod_{\alpha\in H,\alpha>0}  \c 0
\alpha|\vec J, \vec I>
 ;\   \ \ \vec J + \vec I + 2 \vec \rho_H=0\} .
\eqn\mishrelco$$ where $\vec \rho_H$ is half the sum of the positive roots of
$H$.

The \tcm s, as will be shown below, are intimately realted to
gravitational ( and $W_N$ )  models. This correspondence occurs
 once the \co of the theory is taken in the space of
irreducible representations of the $J$ sector \ala. To simplify the
picture  we  present here  only the results the   of the $A_1^{(1)}$
\G model.\refmark\us The  results for the general \tcm\
 cases are  given in refs. [\uss,\usss].
The space of physical states is composed of states built on
$J=J_{r,s}$  where $r$ and $s$ are integers with either $r,s\ge 1$
or $ r<0, s\leq 0$ and
 $2J_{r,s} +1 =r-(s-1)(k+2).$
For each such $J_{r,s}$ there is an infinite set of states with
$I=I_{-r-2lp,s}\ \  G=-2l $ and $I=I_{r-2lp,s}\ \  G= 1-2l$ for every
positive integer
$l$.
 For integer $k$ we have $J=0,..,{k\over
2}$.
Let us now examine   the index
interpretation of the torus partition function.\refmark\GK
 One has to  insert the  values of $\hat L_0 =\L_0
-{1\over k+2} [J(J+1) -I(I+1)]$
and  $\hat J^0_{(tot)}=\jt 0 0- ( J+I+1+)$
into $ Tr[(-)^Gq^{\hat L_0}e^{i\pi\theta \widehat
J^0_{(tot)}} ].$
The end result\refmark\us is
 $$  Tr[(-)^Gq^{\hat L_0}e^{i\pi\theta \widehat
J^0_{(tot)}} ]= 2iq^{-1\over
4(k+2)}e^{-i\pi {\theta\over 2}}M_{k,J}(\tau,\theta)
.\eqn\mishbMTT$$
where
$$M_{k,j}(\tau,\theta) =\sum_{l=-\infty}^{\infty}
q^{(k+2)(l+{j+\half\over (k+2)})^2}sin\{\pi\theta[(k+2)l +{j+\half}]\}
\eqn\mishbM$$
 is the numerator of the character which corresponds to the
highest weight  state $J$.
We have, thus,  rederived using the BRST  cohomology
the path integral results of ref. [\GK], for the torus partition
function.  
  \section { Comparison with string models}
 The main  motivation to study the \tcm\  is  the idea that
 they are closely related to non-critical string models. More
specifically,  we expect a correspondence between the \A    \tGH
models and $W_N$ strings and, in particular, between  the  case of
$G=SL(2)$  and minimal models coupled to gravity.
 Therefore, we would like  to examine  now
 whether one  can  map   the topological coset models  into
string models.
In fact, for reasons that will be clarified shortly,  the
comparison with the gravitational models should be done with the
topological coset models only    after twisting their
energy-momentum tensor.     For  $G=$\S N     the
latter is given by
$$T(z)\rightarrow \t T(z) = T(z)
+\sum_{i=1}^{N-1}\pa\jt {} i (z)\eqn\mishetT$$
For the  $N=2$ case this type of twist in $T(z)$ corresponds to an
addition of the term proportional to
$\omega {\bar J}^{tot}_0+\bar\omega \jt 0 {}$ to the
action of eqn. \mishwzwh, where we use the following expression for
the  curvature $R^2 =\bar\pa\omega +\pa\bar\omega $.
It is easy to realize that a similar modification of the action
arises when one introduces  an holonomy  $\theta^0$ in the
parametrization of the gauge fields namely $A^0 = Tr[T^0 h^{-1}\pa h]
+\theta^0$ and identifies $\theta^0$ with $\omega$. In the   general \S N
case the  holonomies  are in the Kartan sub-algebra  and they
give rise to   a $\theta^i \bar{\jt  i {}}+\bar \theta^i \jt  i {}$ term.

Obviously, since $T(z)$ and $\pa\jt {} i (z)$ are BRST exact so is
$\t T(z)$. Thus, the total Virasoro anomaly  is unchanged. However,
the contribution of each sector to $c$ is modified as follows
$$c_J\rightarrow \t c_J =  c_J-d_GC_Gk \ \  \ \ c_{H^{(I)}}\rightarrow \t
c_{H^{(I)}} =  c_{H^{(I)}}+d_{H^{(I)}}C_{H^{(I)}}(k+C_G+C_{H^{(I)}})
,\eqn\mishtc$$ and the shift in the ghost contribution is given by a
similar expression which can be found
 from the  fact that the  sum of the shifts vanishes.
In what follows we consider,   for simplicity, the case of $G=SL(N,R)$.
The twisted  ghost  sector includes the ghosts of a $W_N$
gravity, namely, a sequence of ghosts with dimensions $(i,1-i)$ for $i=2,...,N$
contributing $\t c_{Wgh}=-2(N-1)[(N+1)^2+N^2] $ to $\t c$.  The rest of the
ghosts are paired with commuting fields of  the same conformal structure
coming from the  $J$ and $I$ sectors. For $N=2$ one finds $\t
c_{Wgh}=-26-2\#_{pairs}$ where there are two pairs in the \SOS\  model
and one in the ${SL(2)\over U(1)}$ case.
  The net matter degrees of
freedom have the following  Virasoro anomaly  $c=\t c_J-\half[\t
c_{(gh)}-\t c_{Wgh}]=(N-1)[(2N^2 +2N +1) -N(N+1)(t+{1\over
t})]$ which is exactly that of a $(p,q)$ minimal $W_N$ matter
sector\refmark\FaLu provided  $t\equiv k+N={p\over q}$.
This was explicitly verified by analyzing the dimensions and contributions
to $\t c$ of the various free fields in the $J$ sector\refmark\uss.
The expression for $c$ reduces to that of the ${p,q}$ minimal model
for $N=2$.

 Next we want to compare the partition function of the \tcm\  to
that of the correspondig $(W)$ string models. For simplicity we
concentrate on the  relation between the
$(p,q)$ minimal
models and the  \G for  $G=SL(2)$ at level  $k= {p\over q}-2$.
The character of the minimal models coupled to gravity is given by the
numerator of the matter character of the minimal models\refmark\LZ
Comparing eqn. \mishbMTT\ to the
numerator of the character of the minimal model, it is clear that a
correspondence may be achieved only provided one takes $\tau=-\half\theta$.
Recall that in the topological coset models we integrate in the path-integral
only over $\theta$ ( and not over $\tau$) and the result is $\tau$
independent.\refmark\GK
 In this case  the numerator of the
character in the minimal model which is proportional to $Tr[(-1)^G q^{\hat
L_0}]$ is mapped into  $Tr[(-1)^G u^{\hat L_0 -\widehat \jt 0 0 }]$ where
 $u=e^{2i\pi\theta}$ in the \G
model. The integration over the moduli parameter of the torus is therefore
replaced by the  integration over the moduli of flat gauge connection.
 To establish this mapping we
 compare the number of states at a given level and ghost number in the
minimal models with the corresponding numbers at the same ghost number and
``twisted level" of the \SOS\ model. The latter are given for $J=J_{r,s}$ by
 $$\eqalign{
 I= I_{-r-2lp,s}\ \ G=-2l \qquad&\hat L_0 -\widehat\jt 0 0=l^2pq
+l(qr-sp) \cr
 I= I_{r-2lp,s}\ \ G=1-2l \qquad&\hat L_0 -\widehat\jt 0 0=l^2pq
-l(qr+sp)  +rs\cr
}.\eqn\mishdL$$
In the minimal models we have states built on vacua
labeled  by the pair $r,s$ with $1\leq r\leq p-1$ and $1\leq s\leq q-1$
with $ps>qr$ which have dimension $h_{r,s} = {(qr-sp)^2-(p-q)^2\over 4pq}$.
The levels of the excitations  are $\hat L_0=\Delta-h_{r,s}$.
  For $G=2l+1$ one has $\Delta=A(l)={[(2pql+qr+sp)^2-(q-p)^2]\over 4pq}$ and
for $G=2l$ $\Delta=B(l)={[(2pql-qr+sp)^2-(q-p)^2]\over 4pq}$.
\refmark{\BerK,\LZ}
Hence,
the  the contribution of the various levels to the partition function
 are identical to those of $\hat L_0 -\widehat\jt 0 0$ in  eqn. \mishdL\ for
 the
same ghost numbers. The respective vacua satisfy $J=\sqrt{p\over 2 q}p_m$
and
$I=-\sqrt{p\over 2 q}p_L$ where $p_m$ and $p_L$ are the matter and Liouville
momenta respectively. It is thus clear that for a given $r,s$ we get the same
number of states with the
same ghost number parity in the two models and  the two partition functions
on the torus are in fact identical.

\section { c=1 string  as a  \tcm }
Now that the connection between the fractional level \tGH  models
and the $W_N$ string models has been established, we would like to
describe the two dimensional string theory, the $c=1$ model, as a
\tcm.  It is straightforward to realize that \SOS model with $k=-1$
or ${SL(2,R)\over U(1)}$ with $k=-3$  conatins a
matter sector with $c=1$. There is however an important difference
between the $c<1$ and $c=1$ cases. In the latter there is no
background charge for the matter sector and thus  no double
complex in the BRST structure. Recall that the latter enabled us
 to use the $(+,-)$ bosonization
inspite  of  the lack of an explicit \S 2 invariant vacuum.  In the $(+,+)$
scheme there is no aparent way to define degrees such that the
spectral sequence machinery can be applied. A direct computation
of the BRST cohomology seems also intracktable due to the cubic
and quartic terms in $Q$.  The new idea  with which enables us to
bypass all these obstacles, is  to similarity
transforam $Q$  into an operator with an isomorphic \co,
The new operator is a sum  of terms
acting on different sectors  so that its cohomology is a direct
sum  of simpler cohomologies.
A detailed analysis of this appraoch is presented in
ref.[\ussss]. Here we give  a brief description of the method and its
results.   In the  $(+,+)$ parametrization  $T$  and  $\jt {} {} $ are
 given by
$$T^{(total)} =-\dpP\dpM -i\pa^2\pP -\bP\pa\gP -\bM\pa\gM
-\rP\d\cM-\rM\d\cP-2\r0\d\c0. \eqn\mishT$$
$$\eqalign{ \jt {} + =& \bP + \cP\rho^0-\chi^0\rP\cr
{J^0}^{(total)} =&\bP\gP+\bM\gM +i\dpM+\cP\rM-\cM\rP\cr
\jt {} - =& \bP({\gP}^2+{\gM}^2)+ \bM\gP\gM -2i(\gP\pM+t\gM\pP)
+2(2\pa \gP -t\pa \gM)+\cM\rho^0-\chi^0\rM.\cr}\eqn\mishjtPP$$
where $t=k+2$.
The $(\bP,\gP)$ and $(\bM,\gM)$ are $(1,0)$ commuting system and
$\pP$ and $\pM$ are scalars.

Let us now define  the  dimension (0,0) operators of zero ghost
number $$\eqalign{R&=\intz(\cP\rP\gM\gM + 2\cP\rz\gP -
\cP\rP\gP\gP)\cr
P&=-\intz'(i\pP(\bP\gP+\bM\gM+\cP\rM-\cM\rP))',\cr}\eqn\mishPR$$
where $\intz'$ means that the zero modes of $\pP$ are excluded. We
then use these operators to transform $\QBr$ to the desired form
in the following way $$e^{-P}e^R \QBr
e^{-R}e^P = \Qt\eqn\mishQpri$$ with
$$ \eqalign{ \Qt &= \intz[\cM\bP+2i\cz\dpM
-2t\cP\d\gM-2t\pP_0\cP\gM]\cr
&= 2\sum_{n\neq 0}\cz_{-n}\pM_n
+\sum_n(\cM_{-n}\bP_n-2t(\pP_0-n-1)\cP_{-n}\gM_n).\cr}\eqn\mishQtr$$
The mode expansions are relative to the vacuum of the
twisted theory (i.e. $\gamma(z) =\sum_n\gamma_n z^{-(n+1)}$).
{}From \mishQpri\ it follows that the cohomologies of $\QBr$ and of $\Qt$
are isomorphic,
namely, for every state $|\Phi_0>$ in the cohomology of $\Qt$,
the state $|\Psi>=e^{-R}e^P|\Phi_0>$ is in the cohomology of $\QBr$ and
vice versa.

  On the following  direct sum of  Fock spaces
$$\bigoplus_{n\neq 0}F(\cz_{-n},\rz_n,\pM_{-n},\pP_n)
  \bigoplus_n F(\cM_{-n},\rP_n,\gP_{-n},\bP_n)
  \bigoplus_n F(\cP_{-n},\rM_n,\bM_{-n},\gM_n)\eqn\mishFo$$
the first term is subjected to the action of the first term in eqn.
\mishQtr, and  similarly for the second and third terms.
  It is thus  apparent that $\Qt$ indeed decomposes into a sum of
anti-commuting terms which act on separate Fock spaces and, therefore,
that the  \co ring  is a direct sum of smaller ones.

The nontrivial $\Qt$-cohomology states are  spanned by
$$|\Phi_0>=\rM_{-n}{\gM_{-n}}^r|\pP_0=-n+1,\ n> 0,\pM_0=r\rangle $$
$$|\Phi_0>={\gM_{-n}}^r|\pP_0=-n+1,n>0,\pM_0=r-1 \rangle $$
$$|\Phi_0>=\cP_{-n}{\bM_{-n}}^r|\pP_0=n+1 > 0,\pM_0=-r-2 \rangle$$
$$|\Phi_0>={\bM_{-n}}^r|\pP_0=n+1 > 0,\pM_0=-r-1 \rangle $$
for $r=0,1,2,...$, and
$$|\Phi_0>=|any \pP_0,\pM_0=-1>.$$

We can now insert $|\Phi_0>$ into the expressions for  the
states in the \co of $\QBr$ as follows
  $$|\Psi>=  \sum_{n=0}^{\infty}{{(-1)^n} \over {n!}}R^n
           \sum_{m=0}^{\infty} {{P_0}^m \over {m!}}|\Phi_0> =
           e^{-R}e^{P_0}|\Phi_0>  $$
\section{ Physical states of the \SOS\ models versus those of the
$c\leq 1$ models.}
The use of the similarity transforamtion method was motivated by
the $k=-1$ \SOS case which corresponds to the $c=1$ string model.
In fact this appraoch is also adequate for any rational vaule of
$t$ and thus produces a unified description of the  \tcm s  which
are the counterparts of  the $c\leq 1$ Liouville models.
At  ghost number $N_G=-1$,
we expect that the discrete states found above would
correspond to elements of the  ground ring\refmark\Wgr
(recall the shift in the ghost number when moving from states to operators
because $|0>_{phys}=\ch 1 + |0>_{SL(2,C)} $). The lowest level
state is simply $\rh {-1} - |\pP_0=\pM_0=0>$ which corresponds to the
identity operator. The next two states  of  the   \co
 of $\Qt$ which are at level 2 translate  into operators
in the \co of $\QBr$  as follows :
$$\eqalign{ \rh {-1} - \gM_{-1}|\pP_0=0,\pM_0=1>& \rightarrow \t x
=\gM e^{i\pP}\cr
  \rh {-2} - |\pP_0=-1,\pM_0=0>&\rightarrow
\t y = [-i\pa \pP + \ch {} + (\rh {} - +2 \rh {} 0 \gP + \rh {}
+[(\gM)^2 - (\gP)^2])]e^{-i\pM}\cr}\eqn\mishxy$$
 These states are (with the
 appropriate identification) at the same momenta
as those of  the ground ring generators in the $c\leq 1$ models.
  In fact $\t y$
is  equal to $y$ of ref. [\Wgr] with some additions from the ``topological
sectors". One can also change
the form of $\t x$ so it resembles that of the
ground ring $x$  by adding a $\QBr$ exact term as follows
$$\t x =\{\QBr, \half \rh {} 0 (\bM)^{-1}e^{i\pP}\} +
({\bM})^{-1}(\ch {} + \rh {} - +i\pa\pM + \bP\gP -\ch {} - \rh {} +
)e^{i\pP}\eqn\mishtx$$    The ground ring cohomology is now
generated by
 $\t x^n \t y^m$.
As in  the ground ring of ref. [\Wgr], it is easy to realize that
 area preserving diffeomorphisms leave the ground ring invariant.  These
$W_\infty$ transformations
are generated by  currents constructed by acting on
the $N_G=1$ \co operators with $G_{-1}$. Recall that $G= \rh {} - (J^+
- I^+) + 2\rh {} 0 (J^0 - I^0) + \rh {} +(J^- - I^-) +  \pa\rh {}
0$. For instance the generators $\pa_{\t x}$  and
 $\pa_{\t y}$  take the following form
$$\eqalign{\pa_{\t x}=& G_{-1}(\ch {} + e^{-i\pP}) = \bM e^{-i\pP} \cr
\pa_{\t y}=& G_{-1}(\ch {} + (\bM)^{-1} e^{i\pM} )= e^{i\pM}
\cr}\eqn\mishdxdy$$
It is easy to check that indeed, as is hinted by  the notations,
$$\pa_{\t x} \t x= \pa_{\t y} \t y= 1\qquad
\pa_{\t x} \t y= \pa_{\t y} \t x=0.\eqn\mishdx$$
 One may wonder about the operator $({\bM})^{-1}$ which
does not seem to be
an appropriate operator to use since  $\bM= J^+-I^+$.
Without the inclusion of arbitrary  powers of $\bM$ the
space of physical states of the \SOS\ model does not recover that
of the $c\leq 1$ models. A similar
situation is facing us also in the tachyonic sector.
One   possible prescription for regaining
a full equivalence in the
states is to  implement an idea of ref. [\MuVa] where a further
bosonization is invoked for  the $(\bM, \gM)$ system. In
this bosonization
$\bM\equiv e^{u-iv} $ and $\gM\equiv -i\partial ve^{-u+iv}$,
where $u,v$ are free
bosons with a background charge of $-\half$ and ${i\over 2}$
respectively. In terms of the latter bosons, one is entitled to
take any arbitrary power of $\bM$ and hence we complete the missing
states in the comparison with the gravitational models.
For another prescription see ref. [\ussss].
One branch of the tachyons of the $c\leq 1$ model can be easily
identified with a sector of the \co of the \SOS\ model: this is the
vacuum of the latter, $|\pP_0=p^+,\pM_0=-1>$ which corresponds to the
operator $\cP e^{ip^+\pM -i\pP}$. If one identifies $\phi_J$ with
the matter field $X$, $\phi_I$ with the Liouville field $\phi$
and $\cP$ with $c$, the tachyonic states of one branch are indeed
found.  However,  the other branch $\cP e^{ip^-\pP +i\pM}$,
is missing in the \co of $\QBr$.  There are, however,
additional states  with no excitations at $N_G=0$. These are the
states $(\bM_0)^r |\pP_0=1,\pM_0 =-r-1>$ corresponding to the
operators $\cP (\bM)^r e^{i\pM_0\pP +i\pM}$.
Apart from the appearence of the operator
$\bM$ these states are identical  to a discrete series of the other
branch of the tachyons.  If again we bosonize $\bM$ then $r$ can
take  any real number  and thus one finds states which
correspond to the full missing branch.
For $k=-1$, restricting the
values of $r$  to the integers would correspond to
the $c=1$ model at the self-dual radius.

The states of other ghost number are also in one to one
correspondence with those of the $c\leq 1$ Fock space relative
cohomology. The only exception is that our second branch of the
tachyon appears in both $N_G=1$ and $N_G=2$ whereas in the Liouville
model it appears only in the former.  A similar  situation is
revealed in the \SOU\ analysis of  ref. [\MuVa].
\refout

\end